\documentclass[fleqn,twoside]{article}
\usepackage{espcrc2}

\usepackage{graphicx} 
\usepackage[figuresright]{rotating}

\newcommand{\AmS}{{\protect\the\textfont2
  A\kern-.1667em\lower.5ex\hbox{M}\kern-.125emS}}

\title{The QCD Phase Diagram at Non-zero Baryon and
Isospin Chemical Potentials 
\thanks{Talk presented by D. Toublan, Lattice 2004, Fermilab}
\thanks{Work supported in part by NSF grant NSF PHY03-04252, and by US
  DOE grant DE-FG-88ER40388.}}

\author{D. Toublan\address[UIUC]{Department of Physics, University of
    Illinois at Urbana-Champaign, Urbana, IL 61801, USA}, 
    B. Klein\address[Heidelberg]{Institute for Theoretical Physics,
      University of Heidelberg,  69120 Heidelberg,
      Germany}
    and  
    J.J.M. Verbaarschot \address[SUNYSB]{Department of Physics and
      Astronomy, SUNY at Stony Brook, Stony Brook, NY 11794, USA}} 
       
\begin{document}

\begin{abstract}
In heavy ion collision experiments as well as in neutron stars, both
baryon and isospin chemical potentials are different from zero. 
In particular, the regime of small isospin chemical potential is 
phenomenologically important.
Using a random matrix model, we find that the phase diagram
at non-zero temperature and  baryon chemical potential is greatly altered
by an arbitrarily small isospin chemical potential: There are two first 
order phase transitions at low temperature, two critical endpoints, 
and two crossovers at high temperature.  
As a consequence, 
in the region of the phase diagram explored by RHIC experiments,
there are {\it two} crossovers that separate the hadronic phase from
the quark-gluon plasma phase at high temperature. 
\vspace{1pc}
\end{abstract}

\maketitle

\section{INTRODUCTION}

Heavy ion collision experiments are important
for our understanding of the strong interaction at nonzero
temperature and density.
In heavy ion collision experiments, both the baryon and the isospin
densities are different from zero. The time
between the formation of the fireball and its freezeout is so short
that only the strong interactions
play a significant role and
baryon number as well as isospin are conserved. 
Therefore, it is phenomenologically worthwhile to study the effects of
a nonzero isospin chemical potential,
$\mu_I$, on the QCD phase diagram at nonzero temperature, $T$, and baryon
chemical potential , $\mu_B$.
However, most studies at high temperature and nonzero density 
have been restricted to cases where either $\mu_I$ or $\mu_B$ is zero. 

For $\mu_B\neq0$ and $\mu_I=0$, the phase diagram is very
rich. At low $T$ and 
high $\mu_B$, the ground state is believed to be a color
superconductor \cite{cscRev}. If $T$ is
increased and $\mu_B$ is decreased, a first order phase
transition separates the hadronic phase from the quark gluon plasma
phase. If $\mu_B$ is further decreased, this first
order critical line stops in a critical endpoint. At lower $\mu_B$, 
there is a
crossover between the hadronic phase and the quark gluon plasma
phase. These results are based on different models
\cite{modelMuB,RMTmuB}, as well as 
on exploratory lattice studies at low chemical potential 
\cite{lattMuB}.

In the case of $\mu_B=0$ and $\mu_I\neq0$, 
the 
fermion determinant is real and traditional
lattice methods can be applied. In this case, the phase diagram is
very rich as well. At low $T$, an increase in $\mu_I$ 
above the pion mass leads to a superfluid phase
with a pion condensate. At low $T$, 
the phase transition between the hadronic
phase and the pion condensation phase is second order and has mean
field critical exponents. If $T$ is increased, this second
order phase transition becomes first order. Therefore there is a
tricritical point in the phase diagram. These results were found using
both lattice simulations \cite{lattMuI} and effective theories
\cite{chptMu}. At 
high $T$ and low $\mu_I$, a crossover separates the
hadronic phase from the quark gluon plasma phase. There also might be
a critical 
endpoint and a first order phase transition at high $T$ when $\mu_I$
is increased \cite{lattMuI}.

We use a Random Matrix model as a
schematic model for QCD to study the phase diagram at nonzero $T$,
$\mu_B$, and $\mu_I$ \cite{qcdMuBMuI_RMT}.  This model has
been previously used to 
study QCD at $\mu_B\!\neq\!0$ and
$\mu_I\!=\!0$ \cite{RMTmuB}. We then analyze  possible
consequences for heavy ion collision experiments.

\section{RANDOM MATRIX MODEL}

Random matrix models were introduced in QCD to describe the
correlations of  low eigenvalues of the Dirac operator
\cite{RMT}. It was shown that for large matrices 
these models are equivalent to the mass
term of the chiral Lagrangian that is uniquely determined by the
symmetry of QCD \cite{chptRMT}. 
Therefore, in the chiral limit, Random
Matrix Theory provides an exact analytical description of the
low-lying Dirac spectrum. The idea is to replace the matrix elements
of the Dirac operator by Gaussian random variables subject only to the
global symmetries of the QCD partition function. The temperature and
chemical potentials enter through external fields structured according
to these symmetries.

For two quark flavors of mass $m_1$ and $m_2$, 
the partition function of our model reads
\begin{eqnarray}
  \label{Zrmt}
\int {\cal D} W \; \prod_{i=1}^2 \det \left(
\begin{array}{cc} m_i & {\cal W}_i \\ -\bar{\cal W}_i & m_i
\end{array} \right)  \; e^{-n G^2 {\rm Tr} W W^\dagger},
\end{eqnarray}
where 
${\cal W}_i\!=\!W+\Omega+\mu_i$, 
and $\bar{\cal W}_i\!=\!W^\dagger+\Omega^\dagger-\mu_i$
are $n \times n$ matrices, with $\Omega\!=\!{\rm diag}(iT,-iT)$. 
The elements of the random matrix $W$ are complex. First, we express the
determinant in (\ref{Zrmt}) as a Grassmann integral. Then, we perform the
Gaussian integration over $W$. Third, the resulting four-fermion
interaction is decoupled by means of a Hubbard-Stratonovich
transformation at the expense of introducing mesonic degrees of
freedom. Finally we perform the Grassmann integration, and
the partition function (\ref{Zrmt}) can 
be mapped onto \cite{qcdMuBMuI_RMT}
\begin{eqnarray}
Z=\int {\cal D}A \; \exp(-{\cal L}(A, A^\dagger)),
\label{Za}
\end{eqnarray}
where  $A$ is an arbitrary complex $2\times2$ matrix, and
\begin{eqnarray}
  {\cal L}=n G^2 {\rm Tr} (A-M^\dagger)( A^\dagger-M) - \frac n2 {\rm
    Tr} \log Q^\dagger Q ,
\label{Zla}
\end{eqnarray}
with $M={\rm diag}(m_1,m_2)$, and 
\begin{eqnarray}
  Q = \left ( \begin{array}{cc}
A  &  iT + \mu_B+\mu_I I_3 \\
iT+\mu_B +\mu_I I_3 & A^\dagger  
\end{array} \right ),
\end{eqnarray}
with $I_3={\rm diag}(1,-1)$. This is an exact mapping.

In the large-$n$ limit, this partition function can be evaluated by a
saddle point approximation. To solve the saddle point equations, we
make an Ansatz for the matrix $A$ based on the symmetry of the partition
function. Since we have two independent
chemical potentials, one for each quark flavor, the chiral
condensates are not necessarily equal. Furthermore, at high enough
$\mu_I$, we expect a pion condensate. 
We thus assume that
\begin{eqnarray}
  A = \left ( \begin{array}{cc}
\sigma_1  & \rho \\
\rho & \sigma_2  
\end{array} \right ).
\end{eqnarray}
In this parameterization, the chiral condensates are given by
$\langle\bar{u}u\rangle\!=\!G^2(\sigma_1-m_1)$,
$\langle\bar{d}d\rangle\!=\!G^2(\sigma_2-m_2)$, and the pion
condensate by $\frac12( \langle\bar{u}\gamma_5d\rangle -
\langle\bar{d}\gamma_5u\rangle )\!=\!G^2 \rho$. We thus get an
effective potential that can be studied in the usual way.

We limit ourselves to the case $m_1\!=\!m_2\!=\!m$. 
We are particularly interested in the phase diagram in the $\mu_B$-$T$ plane
at $\mu_I$ small enough so that the superfluid phase is never
reached, because it corresponds to the conditions of heavy ion
collision experiments. 
The phase diagram in the $\mu_B$-$T$-plane for zero $\mu_I$ has been studied 
in \cite{RMTmuB}. In the chiral limit, the chiral restoration transition 
extends as a second order line from the $\mu_B\!=\!0$ axis, changes order at a 
tricritical point, and intersects the $T\!=\!0$ axis as a line of first order 
transition.
For nonzero quark mass, the first order transition ends in a critical 
point, and the second order transition becomes a crossover. 
\begin{figure}[h]
\vspace{-0.3cm}\hspace{0.5cm}
\includegraphics*[scale=0.3, clip=true, angle=0, draft=false]{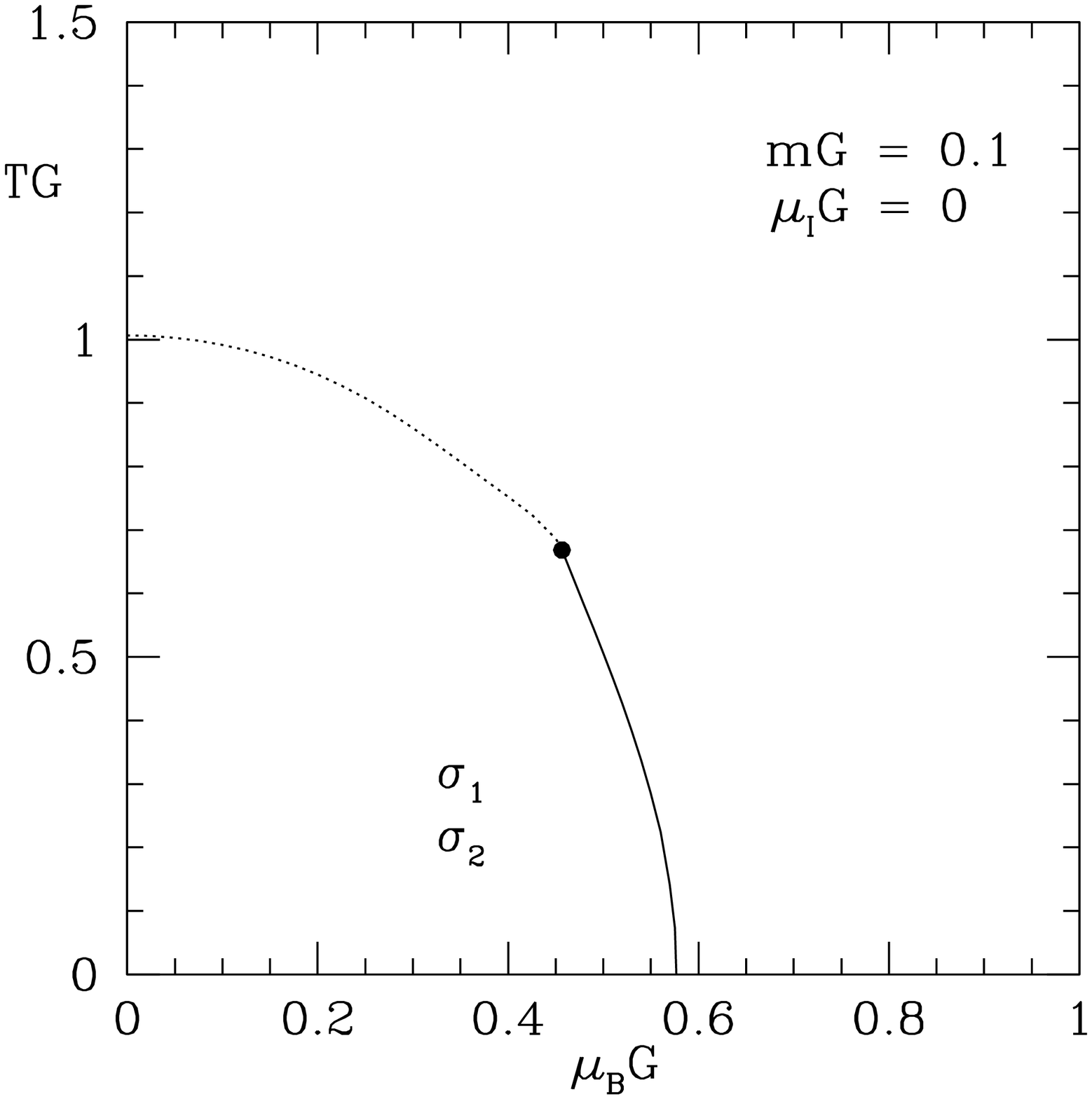}
\vskip 0.2cm \hspace{0.5cm}
\includegraphics*[scale=0.3, clip=true, angle=0, draft=false]{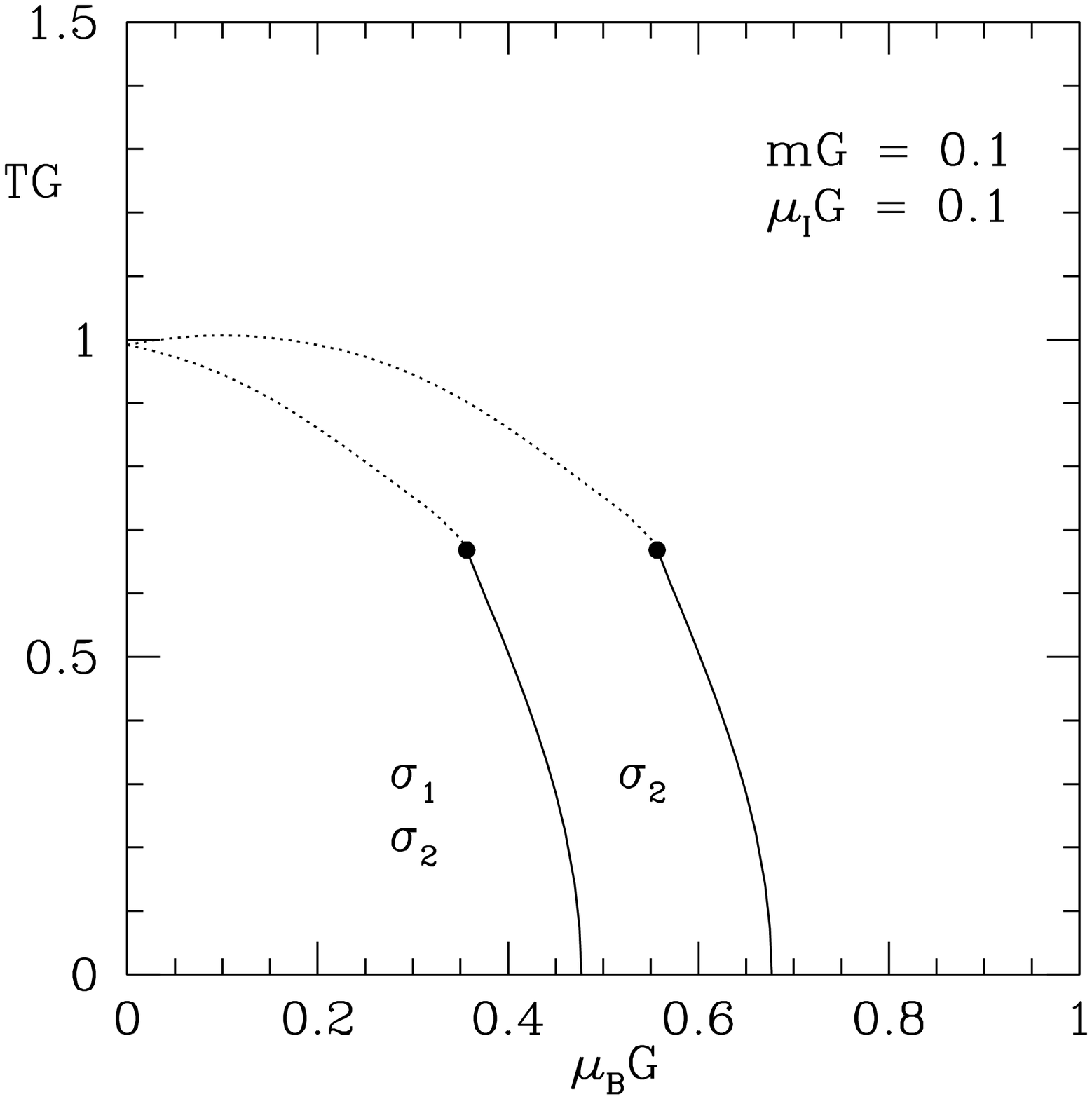}
\vspace{-1cm} \caption{Phase diagram in the $\mu_B$-$T$-plane for
quark mass  $mG\!=\!0.1$ and $\mu_I$ as shown. 
First order transitions are denoted by full curves, and crossovers by
dotted curves. The condensates
$\sigma_1\!=\!\langle\bar{u}u\rangle$ and
$\sigma_2\!=\!\langle\bar{d}d\rangle$ are omitted where $\ll1$.}
\vspace{-0.75cm}
\label{RMTphaseD}
\end{figure} 
Figure~\ref{RMTphaseD} shows the phase diagram in the $\mu_B$-$T$-plane
at finite quark mass $mG\!=\!0.1$ for zero isospin chemical
potential, $\mu_I\!=\!0$, and for  
$\mu_I G\!=\!0.1$. We observe that the first order curve splits into
two first order curves that are separated by $2\mu_IG$. 
This can be understood as follows.  Below 
the threshold for pion condensation,
the free energy separates into a sum over the two flavors.
For $\mu_I\!=\!0$, the chiral phase 
transition lines for both flavors coincide. A nonzero $\mu_I$ 
breaks the flavor symmetry, and the first order transition 
lines for the two flavors split and shift in opposite directions. 
The temperature of the critical endpoints is not affected by
$\mu_I$.

\section{CONCLUSIONS}

We have used a Random Matrix model for QCD at nonzero $T$,
$\mu_B$, and $\mu_I$. 
We have found that in the region of high $T$ and small $\mu_B$, an
arbitrarily small $\mu_I$ greatly alters the phase diagram in the
$\mu_B$-$T$ plane: There are two crossovers, two critical endpoints,
and two first order phase transition 
lines that separate the hadronic phase from the quark gluon plasma
phase \cite{qcdMuBMuI_RMT}. This could have important
consequences for heavy ion 
collision experiments, since they are done at $\mu_B\neq0$ {\it and}
$\mu_I\neq0$. If our Random Matrix model gives an accurate description
of the phase diagram, it might also be interesting to use different
isotopes in heavy ion collision experiments in order to vary $\mu_I$
at constant $\mu_B$. These results have been confirmed by
other models, albeit with some constraints in one case
\cite{qcdMuBMuI_model}. 
Finally, recent lattice studies at small nonzero $\mu_B$
can also be used at nonzero $\mu_B$ {\it and} 
$\mu_I$, and will provide an important test for our results.

\end{document}